\documentclass[a4paper]{article}

\usepackage{graphicx}      
\usepackage{amsmath}       
\usepackage{amsfonts}
\usepackage{amssymb}
\usepackage{titling}
\usepackage{cite}
\usepackage{array}
\usepackage{multirow}
\usepackage{booktabs}      
\usepackage{caption}
\usepackage{subcaption}
\usepackage{threeparttable}
\usepackage{upgreek}
\usepackage{soul}
\usepackage{siunitx}
\usepackage[utf8]{inputenc}
\usepackage[T1]{fontenc}
\usepackage{newtxtext,newtxmath} 
\usepackage[a4paper,margin=1in]{geometry}
\usepackage{multicol}
\usepackage{indentfirst}
\usepackage{setspace}
\usepackage{hyperref}
\hypersetup{colorlinks,linkcolor={red},citecolor={blue},urlcolor={red}}
\usepackage{mwe}           
\usepackage{float}         
\usepackage{authblk}
\usepackage{titlesec}
\usepackage{tabularx}
\usepackage{graphicx}
\usepackage{array}
\usepackage{makecell}
\newcolumntype{C}{>{\centering\arraybackslash}X}
\usepackage{bm}

\titleformat{\section}
{\normalfont\normalsize\bfseries}
{\thesection}{1em}{}

\titleformat{\subsection}
{\normalfont\normalsize\bfseries}
{\thesubsection}{1em}{}

\titleformat{\subsubsection}
{\normalfont\normalsize\bfseries}
{\thesubsubsection}{1em}{}

\titleformat{\paragraph}
{\normalfont\normalsize\bfseries}
{\theparagraph}{1em}{}

\title{{Effect of Source-to-Substrate Distance on the Optical and Electrical Properties of Thermally Deposited CdTe, CdSe and Sb$_2$S$_3$ Thin Films}}

\begin{document}
	\author[1]{Himanshu Sharma Pathok\thanks{Corresponding author:hspathok@gmail.com}}
	\author[2]{Prasanta Kumar Saikia}
	
	\affil[1,2]{Thin Film Laboratory, Department of Physics, Dibrugarh University, Dibrugarh, Assam-786004, India}
	\date{}
	\maketitle
\begin{center}
	\vspace{-55pt}
	*Email:hspathok@gmail.com
\end{center}
\begin{abstract}

In this work, cadmium telluride (CdTe), cadmium selenide (CdSe), and antimony sulfide (Sb\(_2\)S\(_3\)) thin films were deposited using the thermal evaporation technique by varying the source-to-substrate distance (SSD) from 15 to 24 cm (15, 18, 21, and 24 cm). The influence of the SSD on the structural, compositional, optical and electrical properties of the deposited films was systematically investigated. XRD analysis revealed the polycrystalline nature of the CdTe and CdSe thin films, whereas the Sb$_2$S$_3$ thin films exhibited an amorphous nature. EDX spectroscopy confirmed the presence of the constituent elements in the deposited films. Optical analysis revealed a marginal variation in the optical band gap with increasing SSD, while the film thickness, determined using the envelope method, decreased with increasing SSD. The optical band gap energies of CdTe and CdSe were found to be below 2 eV, whereas Sb$_2$S$_3$ exhibited a relatively higher band gap. The relatively low band gap energies and high absorption coefficients of the CdTe and CdSe films make them promising candidates for absorber layer applications in thin film solar cells (TFSCs). The films exhibited relatively low optical transmittance, which is desirable for efficient light absorption in absorber layers. Electrical characterization confirmed the semiconducting nature of all the deposited thin films. Overall, the results demonstrate that the optical and electrical properties of CdTe, CdSe and Sb$_2$S$_3$ thin films can be effectively modified by varying the SSD. The findings indicate that, in addition to the well established CdTe absorber, CdSe thin films show potential for application as absorber materials in TFSCs.
\end{abstract}
\textbf{Keywords:} Thin film; Thermal evaporation; XRD; UV-Vis spectroscopy
\section{Introduction} The continuous increase in global energy demand, coupled with the depletion of conventional fossil fuel resources and growing environmental concerns, has intensified research into sustainable and renewable energy technologies. Among the various renewable energy sources, solar energy is particularly promising because photovoltaic (PV) devices can directly convert sunlight into electrical energy without producing greenhouse gas emissions during operation.

The first generation photovoltaic technology, based primarily on crystalline silicon (c-Si) solar cells (SCs), has achieved high power conversion efficiencies (PCEs) \cite{Moya, Yamamoto}. Conventional c-Si SCs generally involve energy intensive manufacturing processes and relatively high material and production costs. To address these limitations, TFSCs have been developed using thin semiconductor layers deposited on suitable substrates. TFSCs can reduce material consumption and depending on the fabrication approach and device architecture, offer the potential for lower manufacturing costs and large area deposition.

Among the various thin film photovoltaic technologies, CdTe-based SCs have attracted considerable research interest because CdTe possesses a near-optimal band gap for solar energy conversion, a high optical absorption coefficient and suitable electronic properties for photovoltaic applications. These properties enable efficient absorption of incident solar radiation using relatively thin absorber layers, making CdTe an important material for thin film photovoltaic devices.

CdSe is another important II–VI semiconductor material that has attracted considerable attention for thin film photovoltaic applications. CdSe typically exhibits a band gap in the range of approximately 1.71–1.80 eV, depending on factors such as crystal structure, deposition conditions, film thickness and stoichiometry. This band gap enables CdSe to absorb a significant portion of the visible region of the solar spectrum. \cite{Zhao}  O. I. Olusola et al. \cite{Olusola} investigated CdSe thin films and reported their potential application as buffer or window-layer materials in TFSCs. In our previous study, we fabricated CdSe-based TFSCs and obtained a power conversion efficiency of approximately 5\%. These results demonstrate that CdSe can be considered a promising semiconductor for thin film photovoltaic devices. Depending on the device architecture and layer configuration, CdSe has potential for use as an absorber or buffer layer material, although its suitability for each role depends on its optical, structural, and electrical properties as well as the energy-level alignment with the adjacent layers.
Sb$_2$S$_3$ is another important semiconductor material that has attracted considerable attention as an absorber layer for thin-film solar cells (TFSCs) \cite{Myagmarsereejid,Parize}. It possesses a high optical absorption coefficient and a band gap generally reported in the range of approximately 1.10–1.70 eV, depending on factors such as crystal structure, composition, and deposition conditions. These properties make Sb$_2$S$_3$ a promising candidate for thin film photovoltaic applications. Xinnian Liu and co-workers investigated CdS/Sb$_2$S$_3$-based TFSCs and achieved a PCE of 7.66\% \cite{Liu}. Subsequently, several research groups have investigated Sb$_2$S$_3$-based photovoltaic devices using different device architectures and fabrication approaches, with reported PCEs reaching approximately 10.75\%.

Various chemical and physical deposition techniques have been employed for the preparation of CdTe, CdSe, and Sb$_2$S$_3$ thin films \cite{Salem,Wu, Li}. The structural, morphological, optical, and electrical properties of the resulting thin films are strongly influenced by the deposition conditions and process parameters. Therefore, careful optimization of the deposition parameters is essential for obtaining films with the desired properties and reproducible performance. Even relatively small variations in parameters such as substrate temperature, deposition rate, SSD, and deposition time can significantly influence the microstructural and optoelectronic properties of the deposited films.

In the present work, thermal evaporation was employed to deposit CdTe, CdSe, and Sb$_2$S$_3$ thin films. The influence of the source-to-substrate distance (SSD) on the structural, optical, and electrical properties of the deposited thin films was systematically investigated. The variation in SSD was used to examine its effect on film thickness, optical band gap, structural characteristics, and electrical properties, with the aim of identifying suitable deposition conditions for thin-film photovoltaic applications.

\section{Synthesis of the thin films} All the thin films were deposited using a thermal evaporation unit (HINDHIVAC). Analytical-grade materials were used as the source materials. 0.1 gm each high purity CdTe, CdSe, and Sb$_2$S$_3$ powders (99\% purity, SRL) were used for the deposition of the respective thin films. A molybdenum boat was employed as the evaporation source holder. Prior to deposition, the molybdenum boat was properly outgassed by gradually increasing the current under vacuum to remove any residual impurities or adsorbed gases.
\begin{figure}[h]
	\centering
	\includegraphics[width=0.8\textwidth]{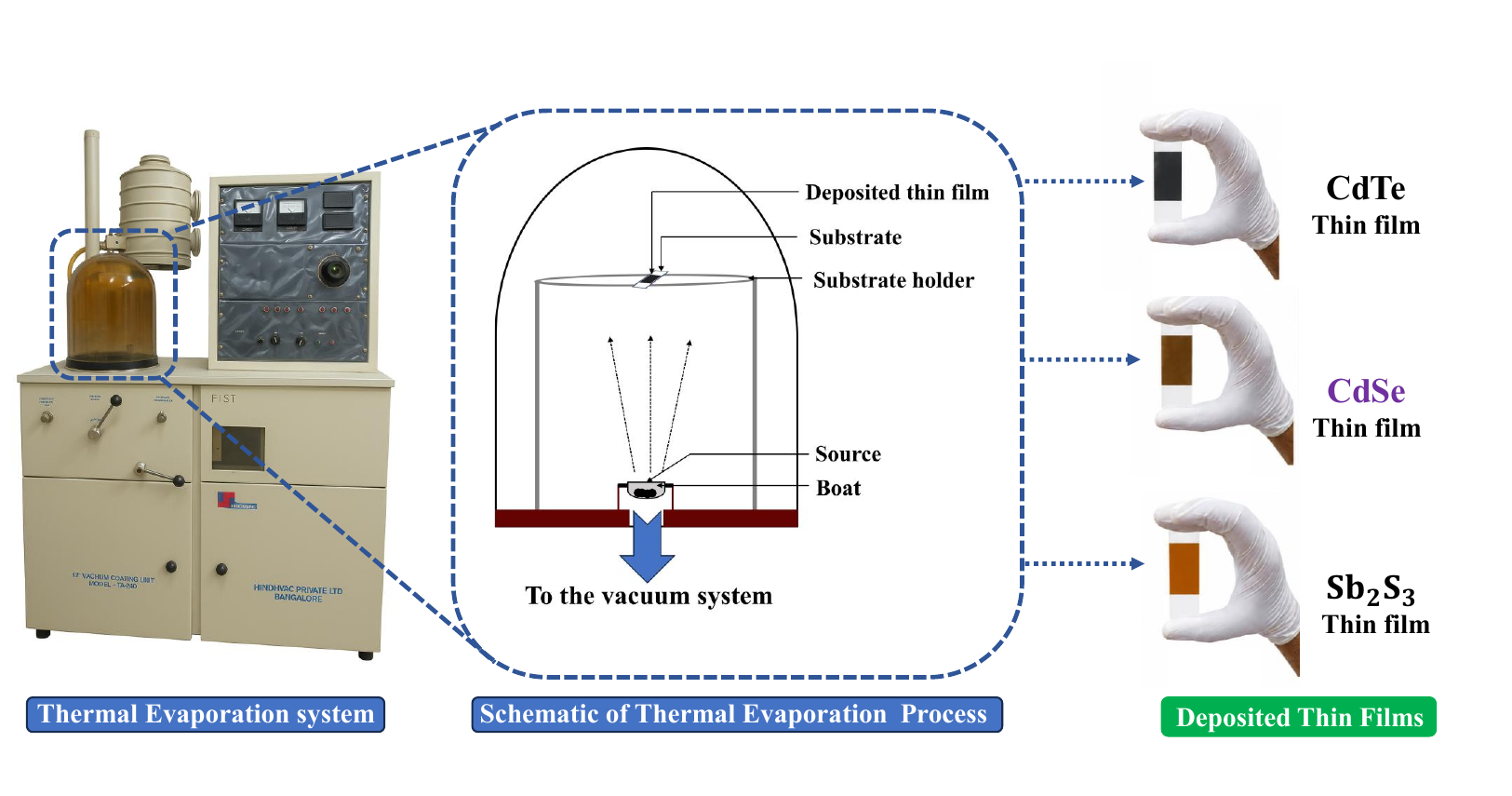}
	\caption{Schematic representation of the deposition process and image of the deposited CdTe, CdSe and Sb\(_2\)S\(_3\) thin films}
	\label{fig:Schematic}
\end{figure} 

Before initiating the deposition process, the evaporation chamber was evacuated to a base pressure of approximately $10^{-6}$ mbar. The current supplied to the molybdenum boat was gradually increased to heat and evaporate the source material at a controlled rate. The evaporated material was transported through the vacuum and condensed onto the glass substrates placed at a predetermined SSD, resulting in the formation of thin films. After complete evaporation of the source material, the current was gradually reduced to avoid thermal shock to the boat and deposited films. The chamber was then allowed to cool under vacuum, and the coated substrates were removed after approximately 24 hours. In table  \ref{tab:Sample},  the sample code of the   CdTe, CdSe ad Sb\(_2\)S\(_3\) are listed.

\subsection{Preparation glass substrate}
For the deposition of thin films, microscopic glass slides with dimensions of \(50 \times 25 \times 0.9~\text{cm}^{-1}\) were used as substrates. Proper cleaning of the glass substrates is essential for achieving uniform and well-adherent thin films. Initially, the glass slides were thoroughly washed with Labolene neutral detergent and distilled water to remove dust and surface impurities. The slides were then immersed in a 10\% nitric acid solution for 4 h to remove oily and grease-like contaminants from the surface. After acid treatment, the glass slides were thoroughly rinsed with detergent and distilled water and subsequently placed in an oven at \(150\,^\circ\mathrm{C}\) for 5 h to remove residual moisture and other surface contaminants. This thermal treatment also helps to minimize surface-related effects that may influence the quality and adhesion of the deposited films. After oven treatment, the glass slides were allowed to cool to room temperature and were subsequently subjected to ultrasonication in deionized water for 15 min. Following ultrasonication, the cleaned glass substrates were considered ready for thin-film deposition.
\begin{table}[h]
	\centering
	\renewcommand{\arraystretch}{1.2}
	\setlength{\tabcolsep}{50 pt}
	\caption{Structural parameters of the CdTe and CdSe thin films}
	\label{tab:Sample}
	\begin{tabularx}{1\textwidth}{ccc}
		\hline
		Thin film & Sample Code & SSD (cm) \\
		\hline
				&	D1	&	15	\\
		CdTe	&	D2	&	18	\\
				&	D3	&	21	\\
				&	D4	&	24	\\
		\hline
				&	S1	&	15	\\
		CdSe	&	S2	&	18	\\
				&	S3	&	21	\\
				&	S4	&	24	\\
		\hline
				&	B1	&	15	\\
		CdTe	&	B2	&	18	\\
				&	B3	&	21	\\
				&	B4	&	24	\\
		\hline
	\end{tabularx}
\end{table}

\section{Characterization} 
The structural, morphological, compositional, optical, and electrical properties of the deposited CdTe, CdSe, and Sb$_2$S$_3$ thin films were investigated using various characterization techniques. The structural properties of the films were analyzed using an X-ray diffractometer (XRD; PANalytical Empyrean) equipped with monochromatic Cu K$\alpha$ radiation. The X-ray tube was operated at a voltage of 40 kV and a current of 30 mA, with a wavelength of $\uplambda=1.5406~\Angstrom$. The surface morphology of the deposited films was examined using field emission scanning electron microscopy (FESEM; ZEISS SIGMA) operated at an accelerating voltage of 20 kV. The elemental composition of the films was investigated using energy-dispersive X-ray spectroscopy (EDX) coupled with the FESEM system to identify and confirm the presence of the constituent elements. The optical properties of the deposited films were studied using a UV-Vis spectrophotometer (Shimadzu UV-2600). The acquired optical data were processed and analyzed using UVProbe software. The electrical properties of the films were evaluated from their current–voltage (I–V) characteristics using a four-probe configuration connected to a Keithley 2450 SourceMeter at room temperature. The electrical measurements and subsequent data analysis were performed using Keithley KickStart software.
\section{Results and discussion}
\subsection{Elemental composition}
\begin{figure}[h]
	\centering
	\includegraphics[width=0.8\textwidth]{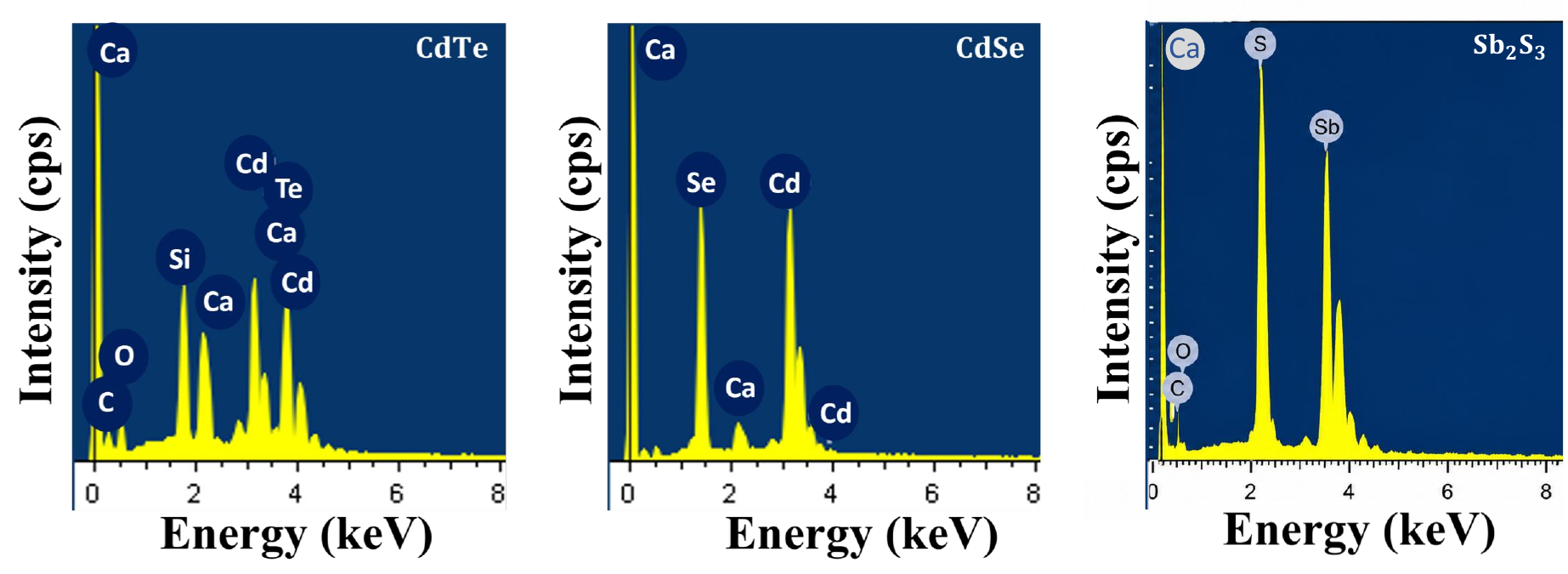}
	\caption{EDX spectra of CdTe, CdSe and Sb\(_2\)S\(_3\) thin films }
	\label{fig:EDX}
\end{figure}

\begin{table}[h]
	\centering
	\renewcommand{\arraystretch}{1.2}
	\setlength{\tabcolsep}{40 pt}
	\caption{EDX spectra  of the CdTe, CdSe and Sb\(_2\)S\(_3\) thin films}
	\label{tab:EDX}
	\begin{tabularx}{1\textwidth}{ccc}
		\hline
		Name of the thin film & Elements & at.\%  \\
		\hline
		CdTe	& C 	& 4.9  	\\
		& O 	& 10.6  \\
		& Si	& 10.5  \\
		& Ca	&  2.2 	\\
		& Cd	&   32.0\\
		& Te	&   39.8\\
		\hline
		CdSe	& O 	&   7.7	\\
		& Se	&   36.2\\
		& Cd	&   53.8\\
		& Ca	&   2.3	\\
		\hline
		Sb\(_2\)S\(_3\) & Sb & 42.6\\
		& O  & 5.8 \\
		& C  & 6.1 \\
		& S  & 43.2\\
		& Ca & 2.3 \\
		\hline
	\end{tabularx}
\end{table}

Figure \ref{fig:EDX} shows the EDX spectra of the CdTe, CdSe, and Sb\(_2\)S\(_3\) thin films (SSD$=$18 cm). The presence of the constituent elements in all the thin films was confirmed by   EDX analysis. The atomic percentages of the constituent elements, after excluding the contributions from Ca and Si originating from the soda-lime glass substrates and O associated with surface moisture/contamination, are listed in Table \ref{tab:EDX}

\subsection{Structural analysis}
\begin{figure}[h]
	\centering
	\includegraphics[width=0.8\textwidth]{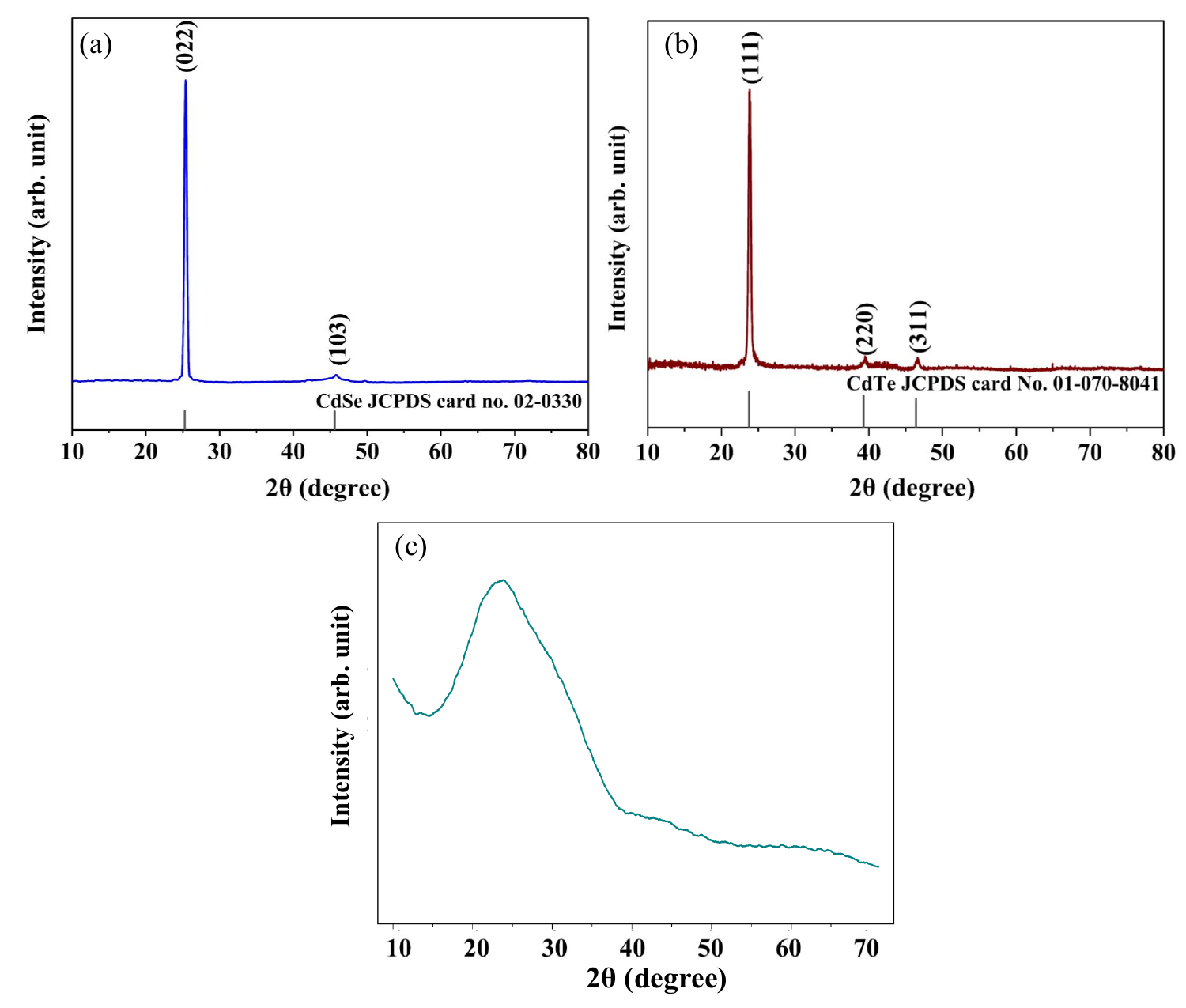}
	\caption{XRD pattern of the (a) CdTe, (b) CdSe and (c) Sb\(_2\)S\(_3\) thin films}
	\label{fig:XRD}
\end{figure} 

\begin{table}[h]
	\centering
	\renewcommand{\arraystretch}{1.2}
	\setlength{\tabcolsep}{40 pt}
	\caption{Structural parameters of the CdTe and CdSe thin films}
	\label{tab:structural}
	\begin{tabularx}{1\textwidth}{cccc}
		\hline
		Sample & 2\(\uptheta\) (degree) & (hkl) & D (nm) \\
		\hline
		CdTe	& 23.83 & (111) & \\
		& 39.50 & (220) & 21.23 \\
		& 46.59 & (311) & \\
		CdSe	& 25.37 & (022) & 19.90\\
		& 45.76 & (103) & \\
		\hline
	\end{tabularx}
\end{table}

Figure \ref{fig:XRD} presents the XRD patterns of the CdTe, CdSe, and Sb\(_2\)S\(_3\) thin films (SSD$=$18cm). The CdTe and CdSe thin films exhibit polycrystalline nature, whereas the Sb\(_2\)S\(_3\) thin film shows an amorphous nature. Comparison of the observed diffraction peaks with the corresponding standard JCPDS data confirms that CdTe crystallizes in the cubic structure (JCPDS Card No.: 01-070-8041), while CdSe exhibits a hexagonal crystal structure. (JCPDS Card No. 02-0330) The structural parameters of the CdTe and CdSe thin films are summarized in Table \ref{tab:structural}. 

The crystallite size (D) was estimated using Eq. \ref{eq:crystallite_size} , and the obtained values confirm the nanocrystalline nature of the CdTe and CdSe thin films.
\begin{equation}
	\mathrm{D=\frac{k\uplambda}{\upbeta\cos\uptheta}}
	\label{eq:crystallite_size}
\end{equation}
Where k is constant, known as shape factor (k=0.94) and \(\uplambda\) is the wavelength of the Cu-k\(\upalpha\) (\(\uplambda=1.5406~\Angstrom\))
Calculated value of the D are listed in Table \ref{tab:structural}

\subsection{Optical analysis}
\begin{figure}[h]
	\centering
	\includegraphics[width=0.8\textwidth]{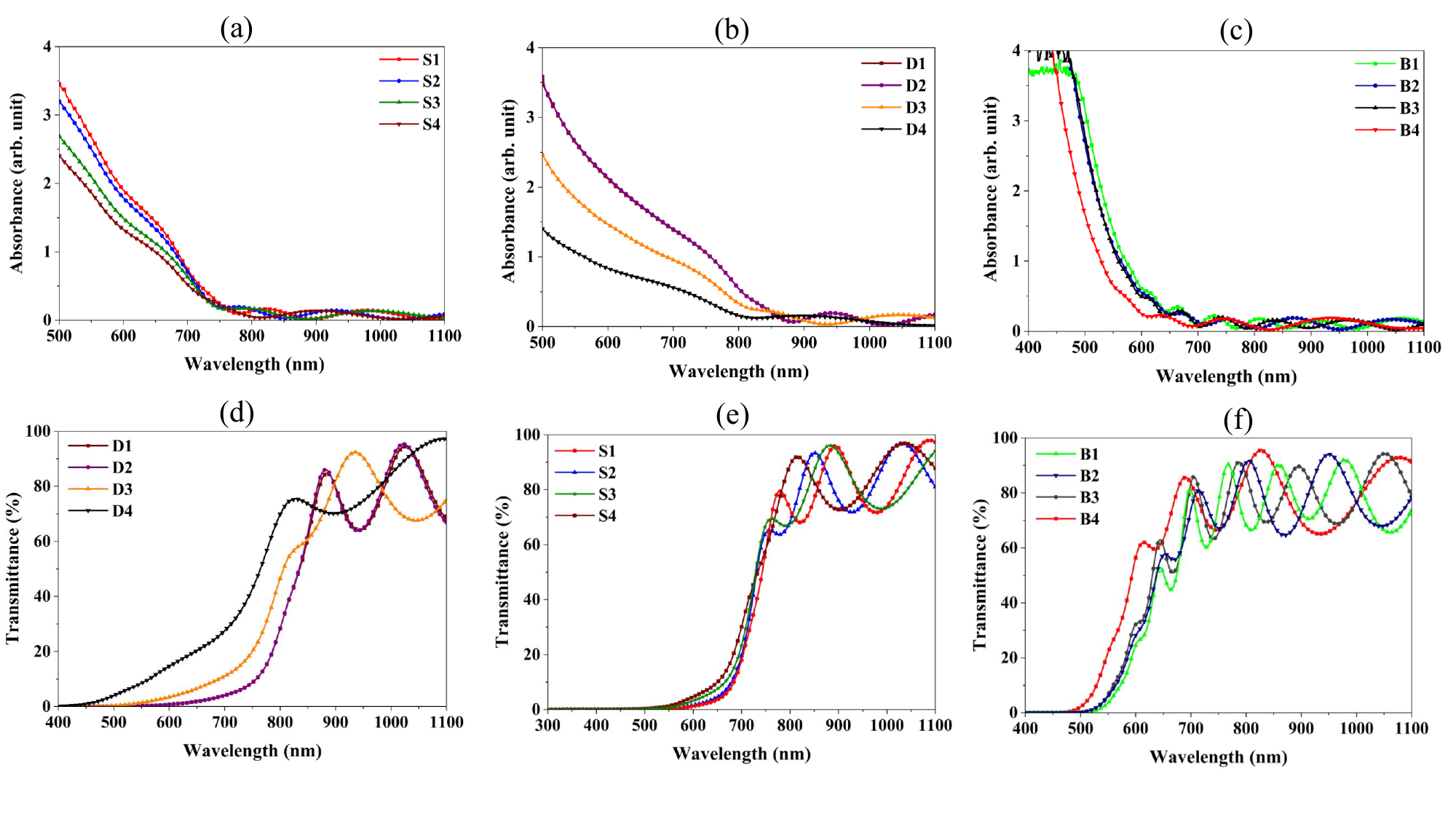}
	\caption{Absorbance versus wavelength spectra of (a) CdTe, (b) CdSe and (c) Sb\(_2\)S\(_3\) thin films and Transmittance versus waavelength spectra of (d) CdTe, (e) CdSe and (f) Sb\(_2\)S\(_3\) thin films }
	\label{fig:Abs_trans}
\end{figure}

Figure \ref{fig:Abs_trans} presents the transmittance and absorbance spectra of the CdTe, CdSe and Sb$_2$S$_3$ thin films deposited at different SSDs. With an increase in SSD, the absorption edge of the thin films shifted towards shorter wavelengths (higher photon energy), indicating an increase in the optical band gap ({E$_\text g$}). The deposited films exhibited strong optical absorption in the visible and near-infrared (NIR) regions, accompanied by relatively low transmittance in the corresponding wavelength range. 

\begin{figure}[h]
	\centering
	\includegraphics[width=0.8\textwidth]{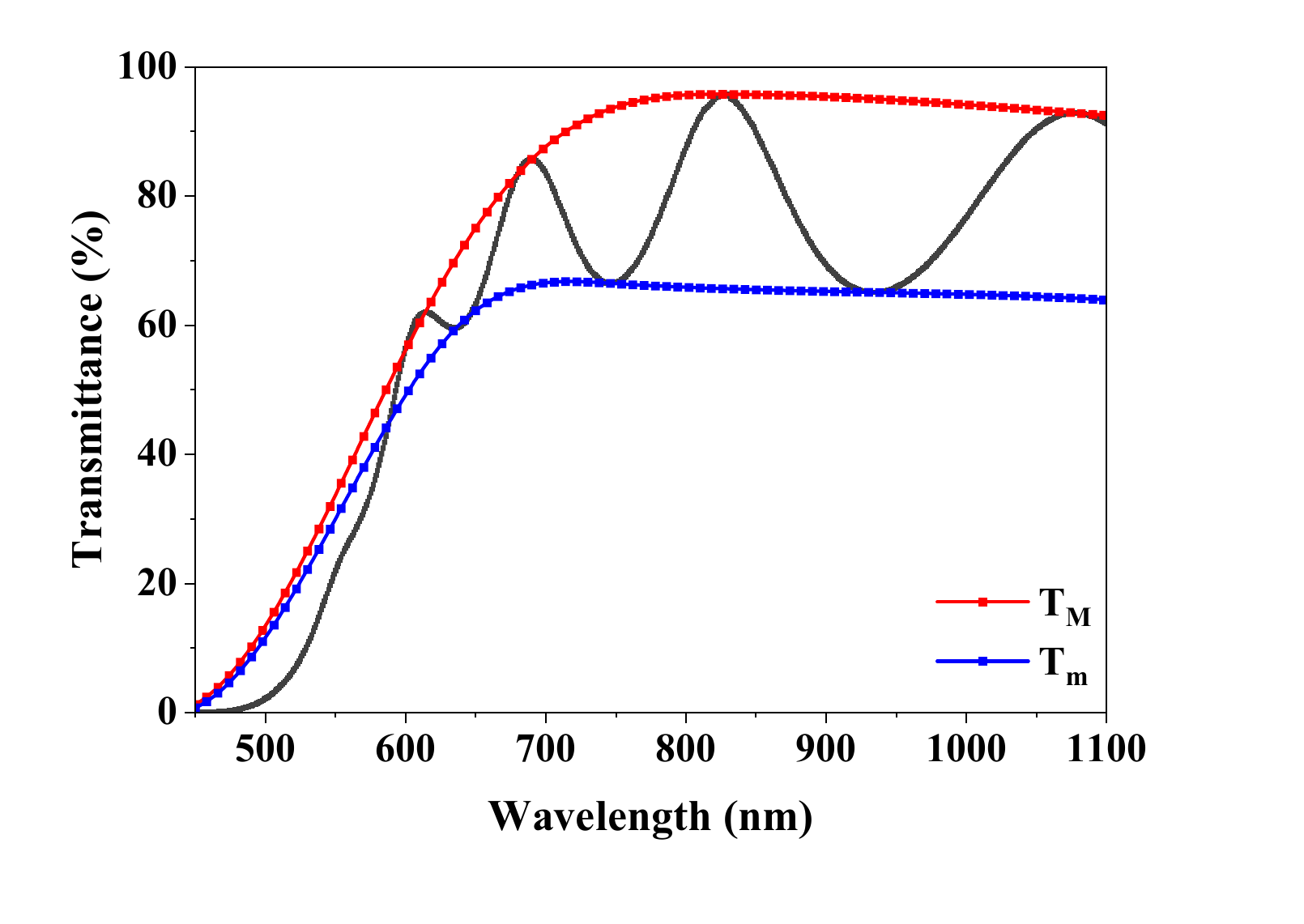}
	\caption{Image for the calculation of T$_\text M$ and T$_\text m$ from the transmittance plot}
	\label{fig:Envelop_method}
\end{figure}

The Swanepoel envelop method (SWEM) enables the determination of the refractive index and film thickness from the interference fringes observed in the transmission spectra of relatively uniform thin films. The refractive index (n) of the thin films was determined using Eq. \ref{eq:R.I.} \cite{Ahmed}.
\begin{equation}
	\text{n}^2=\left[\text N+(\text N^2-\text S^2)^{\frac{1}{2}}\right]
	\label{eq:R.I.}
\end{equation}
Here,
\begin{equation}
	\text N=\mathrm{2 S\left(\frac{T_M-T_m}{T_M.T_m}+\frac{S^2+1}{2}\right)}
\end{equation}
where T$_\text M$ and T$_\text m$ denote the maximum and minimum transmittance values, respectively, at a given wavelength. S represents the refractive index of the glass substrate, with S = 1.51. As illustrated in Figure \ref{fig:Envelop_method}, the T$_\text M$ and T$_\text m$  values of the thin films were determined from the transmittance spectra using Origin software (Origin 2026b).

The thickness (t) of the thin films was estimated using Eq. \ref{eq:thickness} \cite{Hassanien}.
\begin{equation}
	\mathrm{t=\frac{\uplambda_1.\uplambda_2}{2(n_2\uplambda_1-n_1\uplambda_2)}}
	\label{eq:thickness}
\end{equation}
 Where $\uplambda_1$ and $\uplambda_2$ are the wavelength of the two corresponding maxima or minima with refractive index n$_1$ and n$_2$, respectively. The calculated values of the thickness are listed in Table 1.
 
\begin{figure}[h]
	\centering
	\includegraphics[width=1\textwidth]{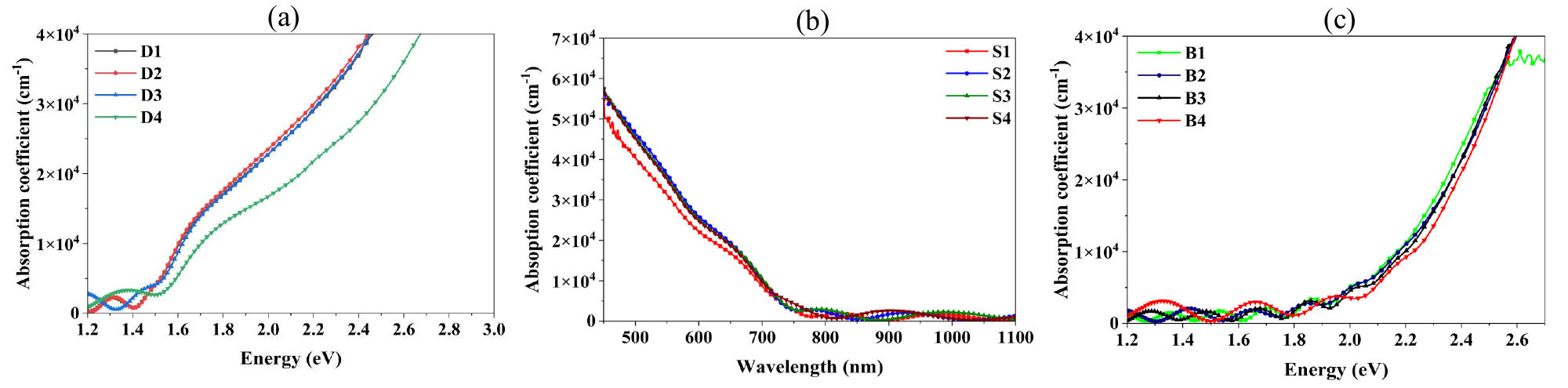}
	\caption{Absorption coefficient versus wavelength spectra of (a) CdTe, (b) CdSe and (c) Sb\(_2\)S\(_3\) thin films}
	\label{fig:Absorption_coefficient}
\end{figure}

The absorption coefficient ($\upalpha$) of the thin films was calculated using Eq. \ref{eq:Absorption_coefficient}. 
\begin{equation}
	\upalpha=2.303\frac{\text A}{\text t}
	\label{eq:Absorption_coefficient}
\end{equation}
where `t' is the thickness of the thin film in cm. The calculated values of $\upalpha$ were found to be in the range of $10^{4}$–$10^{5}$ cm$^{-1}$ in the visible and NIR regions, as shown in Figure \ref{fig:Absorption_coefficient}. The high absorption coefficient observed for these films indicates their strong light absorbing capability. 

The optical band gap (E$_\text g$) of the deposited thin films was estimated using Tauc’s relation (Eq. \ref{eq:Tauc}). 
\begin{equation}
	\left(\upalpha\mathrm{h}\upnu\right)^m=A\left(\text h \upnu-\mathrm{E_g}\right)
	\label{eq:Tauc}
\end{equation}
For $\text m=2$, the Tauc plots exhibited a distinct linear region near the absorption edge, indicating that the films possess a direct allowed band gap transition.  Here A is a proportionality constant. The estimated E$_\text g$ values were found to increase gradually with increasing SSD.
\begin{figure}[h]
	\centering
	\includegraphics[width=1\textwidth]{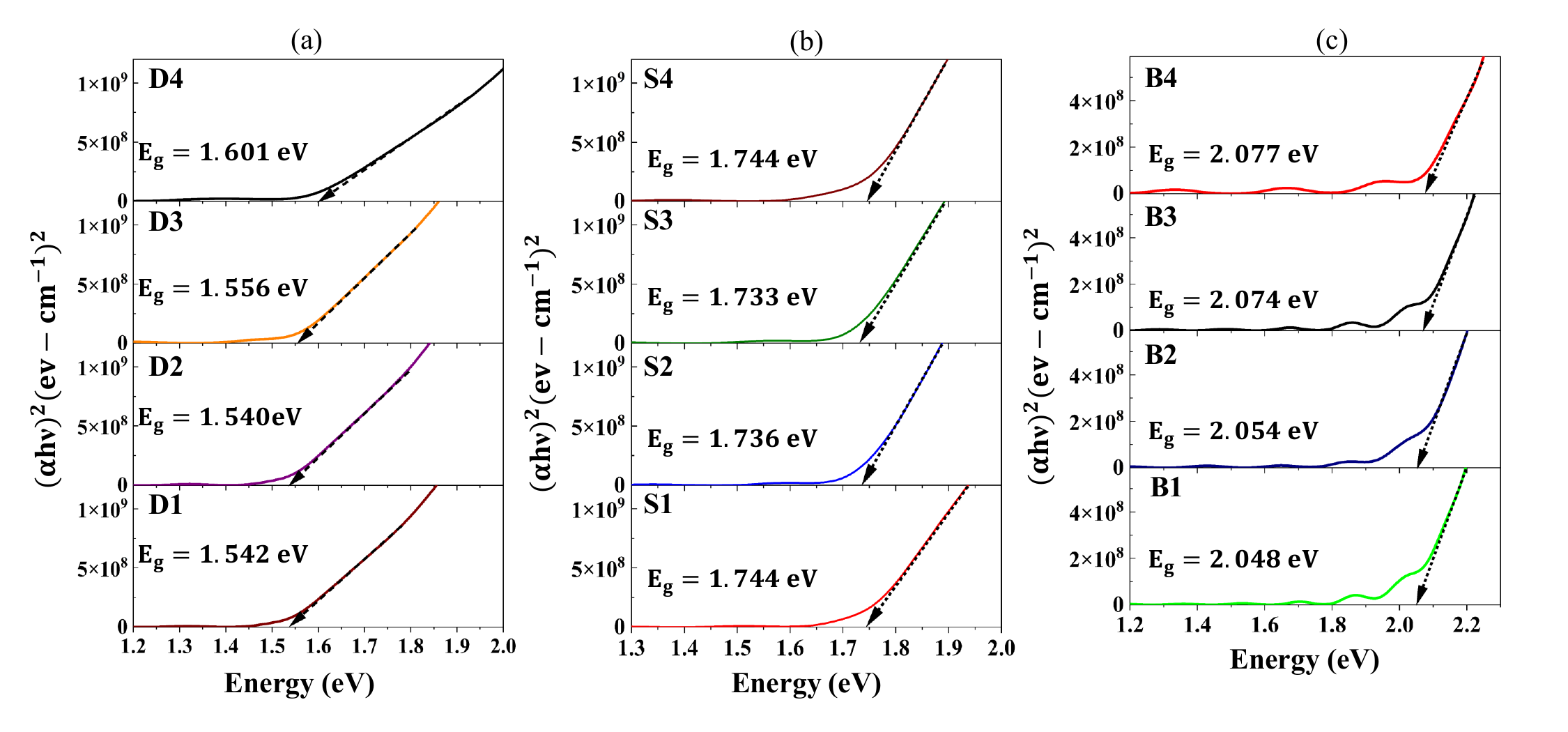}
	\caption{\((\upalpha\text h\upnu)^2\) versus energy plot of (a) CdTe, (b) CdSe and (c) Sb\(_2\)S\(_3\) thin films}
	\label{fig:band_gap}
\end{figure} 

This increase may be attributed to the quantum confinement effect associated with variations in the structural characteristics of the films. The observed blue shift in the absorption edge further supports the increase in the optical band gap. The calculated E$_\text g$ values are presented in Table 1.
\subsection{Electrical analysis}
\begin{figure}[h]
	\centering
	\includegraphics[width=0.7\textwidth]{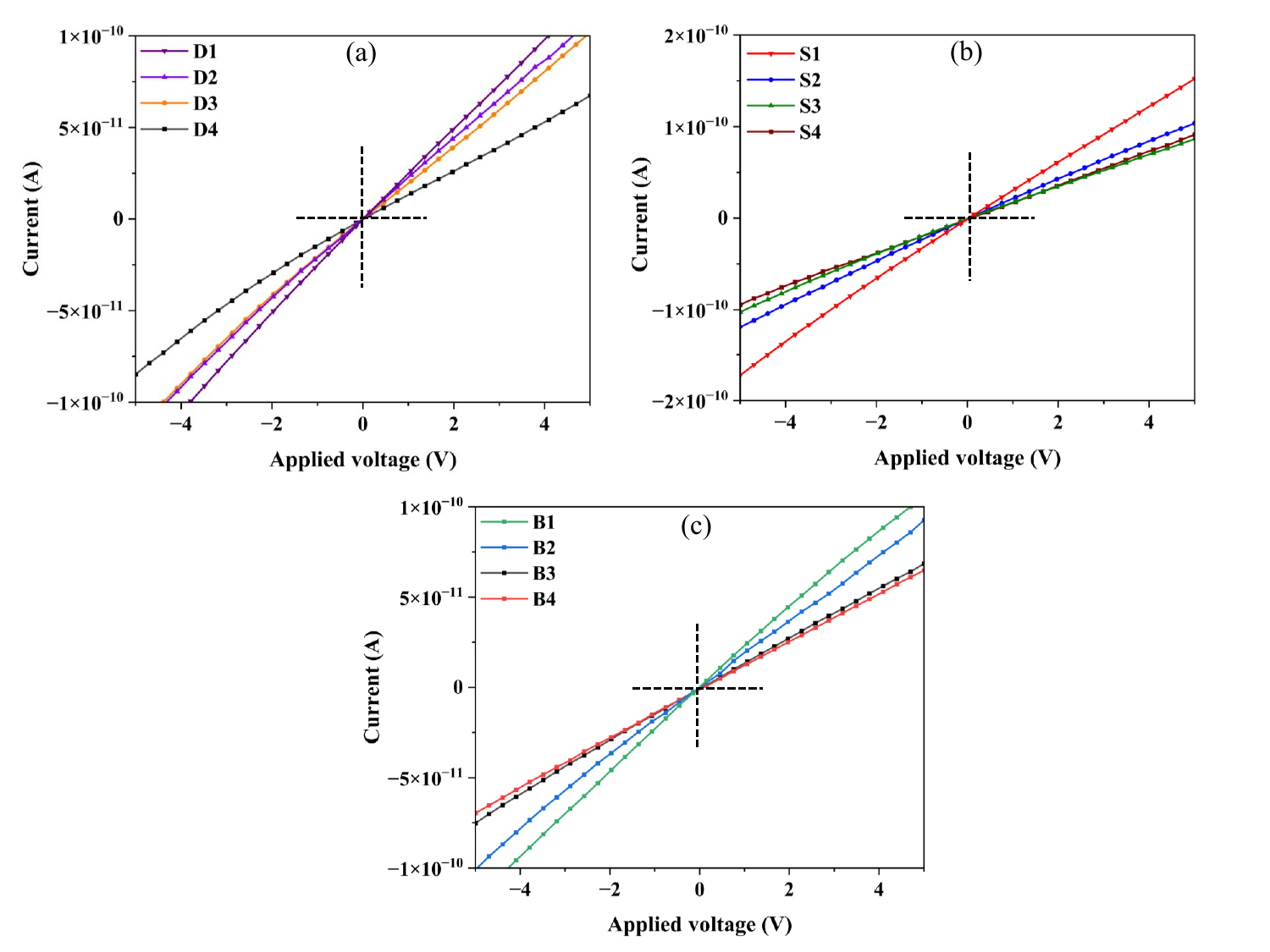}
	\caption{Current versus Applied voltage plot of (a) CdTe, (b) CdSe and (c) Sb$_2$S$_3$ thin films}
	\label{fig:I-V}
\end{figure}

The current–voltage (I$-$V) characteristics of the CdTe, CdSe, and Sb\(_2\)S\(_3\) thin films are shown in Figure \ref{fig:I-V}. The measurements were performed at room temperature under dark conditions. All the investigated thin films exhibited a linear I$-$V response, indicating ohmic conduction behaviour. This linearity suggests a negligible Schottky barrier at the interface between the metal electrodes and the semiconductor thin films. The resistance (R) of each thin film was determined from the inverse of the slope of the linear I$-$V curve.

The resistivity ($\uprho$) and conductivity ($\upsigma$) of the thin films were calculated using Eqs. \ref{eq:resitivity} and \ref{eq:conductivity}, respectively. 
\begin{equation}
	\uprho=\frac{\uppi}{\ln 2}\frac{\text V}{\text I}\text t
	\label{eq:resitivity}
\end{equation}

and, 
\begin{equation}
	\upsigma=1/\uprho
	\label{eq:conductivity}
\end{equation}
The resistivities of all the investigated thin films were found to be of the order of \(10^{7}\ \Omega\cdot\text{cm}\) (Table \ref{tab:Electrical}), while their electrical conductivities were of the order of \(10^{-8}\ \text{S}\cdot\text{cm}^{-1}\), which is consistent with the semiconducting nature of the films.

\begin{figure}[h]
	\centering
	\includegraphics[width=0.8\textwidth]{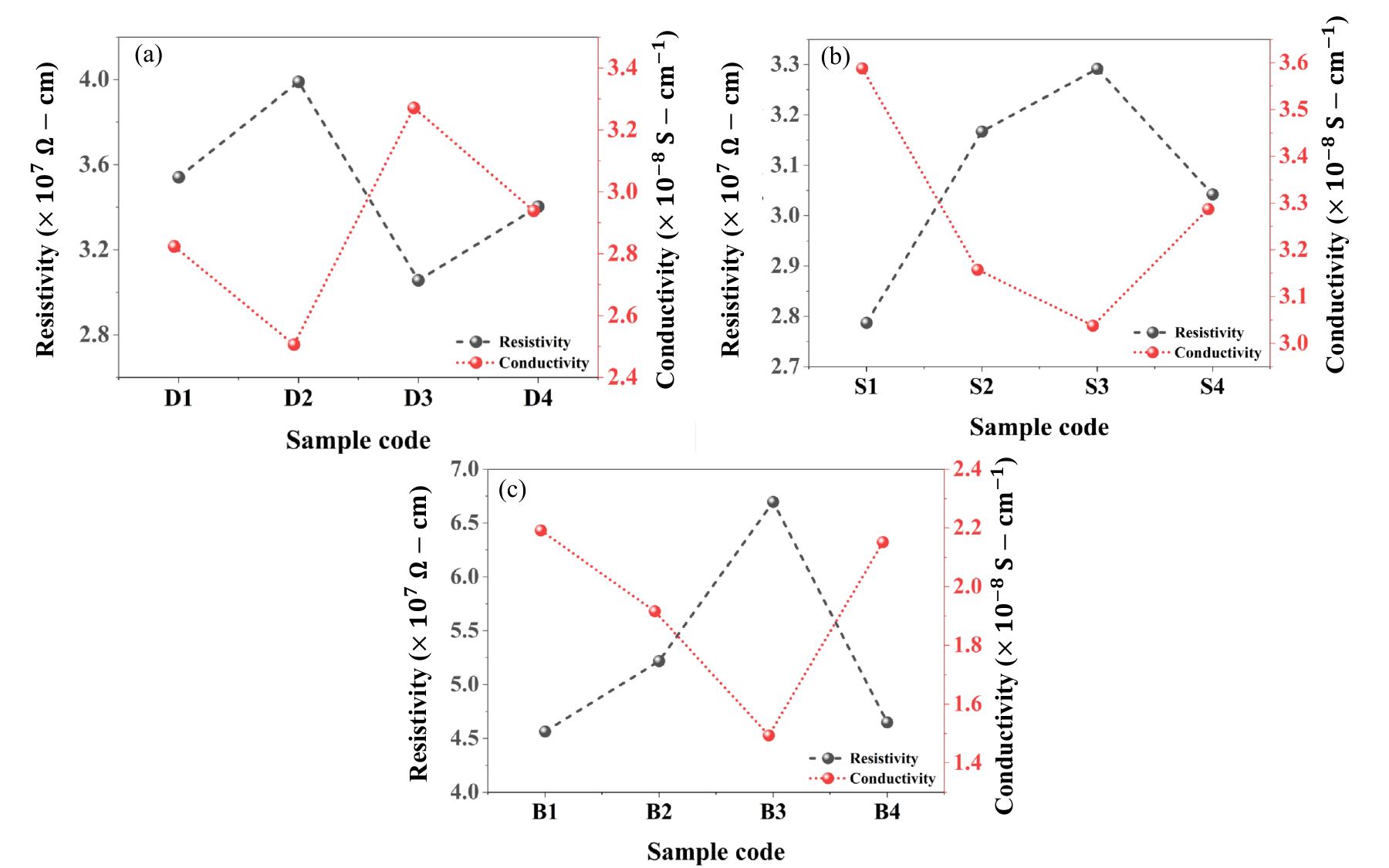}
	\caption{Resistivity and conductivity versus sample code plots of (a) CdTe, (b) CdSe and (c) Sb$_2$S$_3$ thin films}
	\label{fig:R_C}
\end{figure} 
The variations in resistivity and electrical conductivity with sample code are presented in Figure~\ref{fig:R_C}. Among the CdTe thin films, sample D3 exhibits the highest electrical conductivity, whereas sample S1 shows the highest conductivity among the CdSe thin films. For the Sb\(_2\)S\(_3\) thin films, sample B1 exhibits the highest electrical conductivity. These results indicate that the electrical conductivity of the thin films is strongly influenced by the  SSD.

\begin{table}[h]
	\centering
	\renewcommand{\arraystretch}{1.4}
	\setlength{\tabcolsep} {20pt}
	\caption{Electrical parameters of the CdTe and CdSe thin films}
	\label{tab:Electrical}
	\begin{tabularx}{1\textwidth}{cccc}
		\hline
		Sample Code & Resistance & Resistivity (\(\upOmega-\)cm) & Conductivity (S$-$cm$^{-1}$)\\
		\hline
	D1 & 3.97\(\times 10^{10}\) & 3.54 \(\times 10^7\) &  2.82 \(\times 10^{-8}\) \\
	D2 & 4.59\(\times 10^{10}\) & 3.99 \(\times 10^7\) &  2.51 \(\times 10^{-8}\) \\
	D4 & 4.95\(\times 10^{10}\) & 3.06 \(\times 10^7\) &  3.27 \(\times 10^{-8}\) \\
	D4 & 7.09\(\times 10^{10}\) & 3.40 \(\times 10^7\) &  2.94 \(\times 10^{-8}\) \\
	\hline
	S1 & 3.12\(\times 10^{10}\) & 2.79 \(\times 10^7\) &  3.59 \(\times 10^{-8}\) \\
	S2 & 4.41\(\times 10^{10}\) & 3.17 \(\times 10^7\) &  3.16 \(\times 10^{-8}\) \\
	S3 & 5.32\(\times 10^{10}\) & 3.29 \(\times 10^7\) &  3.04 \(\times 10^{-8}\) \\
	S4 & 5.46\(\times 10^{10}\) & 3.04 \(\times 10^7\) &  3.29 \(\times 10^{-8}\) \\
	\hline
	B1 & 4.33\(\times 10^{10}\) & 4.56 \(\times 10^7\) &  2.19 \(\times 10^{-8}\) \\
	B2 & 5.38\(\times 10^{10}\) & 5.22 \(\times 10^7\) &  1.92 \(\times 10^{-8}\) \\
	B3 & 7.04\(\times 10^{10}\) & 6.70 \(\times 10^7\) &  1.49 \(\times 10^{-8}\) \\
	B4 & 7.46\(\times 10^{10}\) & 4.65 \(\times 10^7\) &  2.15 \(\times 10^{-8}\) \\
	\hline
	\end{tabularx}
\end{table}
\section{Conclusion}

CdTe, CdSe, and Sb$_2$S$_3$ thin films were successfully deposited by thermal evaporation at different  SSD and their structural, compositional, optical,and electrical properties were investigated.  XRD analysis revealed that the CdTe and CdSe thin films exhibited polycrystalline nature with cubic and hexagonal crystal structures, respectively, while the Sb$_2$S$_3$ thin films were amorphous. The nanometer-scale crystallite sizes obtained for the crystalline films confirmed their  nanocrystalline nature. EDX analysis confirmed the presence of the constituent elements of the deposited compounds. Optical analysis showed a systematic shift of the absorption edge towards shorter wavelengths with increasing  SSD, accompanied by an increase in the optical band gap. The electrical measurements exhibited linear   I$-$V characteristics, indicating ohmic conduction behaviour, while the measured resistivity and conductivity values were consistent with the semiconducting nature of the films. Furthermore, the electrical conductivity showed a clear dependence on the SSD. These results demonstrate that the SSD is an important deposition parameter for controlling the structural, optical  and electrical properties of thermally evaporated CdTe, CdSe and Sb$_2$S$_3$ thin films.

\end{document}